# A practical theoretical model for Ge-like epitaxial diodes: I. The I-V characteristics

Matthew A. Mircovich,[1] John Kouvetakis[2], and José Menéndez[1,a]

[1]*Arizona State University, Department of Physics, Tempe, AZ, USA 85287-1504*
[2]*Arizona State University, School of Molecular Sciences, Tempe, AZ, USA 85287-1604*

([a]) Electronic mail: jose.menendez@asu.edu

A practical quantitative model is presented to account for the I-V characteristics of *pin* diodes based on epitaxial Ge-like materials. The model can be used to quantify how the different material properties and recombination mechanisms affect the diode performance. The importance of dislocations, non-passivated defects, and residual intrinsic layer doping in determining the qualitative shape of the I-V curves is discussed in detail. Examples are shown covering literature diodes as well as diodes fabricated with the purpose of validating the theoretical effort.

## I. INTRODUCTION

Intense efforts have been devoted since the 1990's to develop technologies that enable the growth of low-defect Ge on Si. (Refs. 1-3). These advances were followed by the growth and characterization of several generations of devices.[4-7] In more recent times, the field has been enriched and expanded by the introduction of GeSn and GeSiSn alloys[8-10] that extend the detection wavelength range and can reach a direct-gap regime.

The basic component of most of the above devices are Ge-on-Si *pn* or *pin* diodes. The optical and electrical characterization of these devices provide important clues about the material quality and the potential of the technology to contribute to the general area of silicon photonics.[7, 11-16] Modern Technology Computer-Aided Design (TCAD) tools can simulate these diodes in great detail, but quantitative fits of experimental I-V or responsivity curves are rare. Instead, the analysis of the experimental data is usually based on fits with functions that contain empirical parameters (such as ideality factors or collection efficiencies) or fits with analytical expressions derived from the depletion approximation. Activation energy studies from measurements of the temperature dependence of the I-V curves are also quite common. This methodology is not ideal because it is very hard to relate empirical parameters to microscopic properties, and because the interpretation of the data is based on assumptions that are difficult to confirm without explicit modeling. On the other hand, the limited use of TCAD tools to fit experimental data reveals the need for alternatives that allow the user to easily match experimental and theoretical I-V curves, even if these alternatives are more limited in scope compared to those included in commercial packages.

In this paper, we present a model for the I-V characteristics of Ge-on-Si diodes that is rigorous enough to allow fitting of experimental data but can be implemented using software and hardware resources commonly available to experimentalists for routine analysis and visualization of experimental data. The model relies on a full numerical solution of the standard semiconductor transport equations. The inherent slowness of this approach is mitigated by the fact that the computation of an I-V curve lends itself naturally to multithreading, and therefore we can produce enough I-V data points very fast. Furthermore, since different material properties affect the experimental data in characteristic ways, the adjustment of the calculated I-V curves to experimental data is greatly facilitated. In this first introduction of the model, we restrict ourselves to homostructure Ge-on-Si *pin* diodes. In subsequent publications we will discuss extensions to GeSn alloys and to general heterostructure diodes. We will also show how the model can be combined with optical codes to generate realistic predictions of diode responsivities.

The numerical solution of the semiconductor equations allows us to assess the validity of simplified approaches. In some cases, such as the use of activation energies to identify recombination-generation mechanisms, we find that the conclusions can be erroneous if the temperature dependence of the band structure and electrical properties are not properly accounted for. In this respect, the model is useful to identify the fundamental electronic and transport properties that still need further research to feed accurate models of Ge-on-Si diodes. Our model also demonstrates that in Ge-on-Si *pin* diodes any residual doping in the nominally intrinsic layer interacts with Trap Assisted Tunneling (TAT) mechanisms to affect the voltage-dependence of the reverse bias current. Simulations under these conditions require an accurate knowledge of the electric field in the structure, which cannot be easily obtained from analytical approximations and requires a numerical solution of Poisson's equation.

We show that a relatively simple description of dislocation-related recombination, consisting of a mid-gap and a shallow Shockley-Read-Hall (SRH) recombination center, is sufficient to reproduce the I-V characteristics of Ge-on-Si diodes grown by different methods, including the standard two-step (low temperature/high temperature) approach for Ge on Si,[4] deposition on germanium-on-insulator (GOI) platforms,[14] and low-temperature approaches based on advanced Ge precursors, as practiced by our group.[17]

The remainder of the paper is organized as follows. In Section II we introduce the model and details on the numerical implementation. In Section III we show how the different model parameters have qualitatively different effects on the experimental I-V curves, which facilitates fitting of experimental data. In Section IV we discuss fits to a few selected examples of Ge-on-Si diodes, and in Section V we summarize our conclusions. A detailed discussion of our material parameter choices is given in an Appendix.

## II. MODEL DESCRIPTION

### A. Basic equations

We will consider a one-dimensional model of a *pin* diode for a two-band semiconductor. We place the origin of the coordinate system at the center of the intrinsic layer, which extends from $z = -d/2$ to $z = d/2$. We assume an acceptor density $N_a(z)$ over a length $W_p$ on one side of the intrinsic layer and a donor density $N_d(z)$ over a length $W_n$ on the other side. In most cases, the acceptor and donor concentrations can be approximated as simple box-like distributions with constant values $N_a$ and $N_d$. The Poisson and current continuity equations for this system are[18]

$$-\frac{d}{dz}\left[\epsilon(z)\frac{dV}{dz}\right] = 4\pi e[p(z) - n(z) + N_d(z) - N_a(z)] \quad (1)$$

$$\frac{1}{e}\frac{\partial j_n(z)}{\partial z} = R(z) \quad (2)$$

$$\frac{1}{e}\frac{\partial j_p(z)}{\partial z} = -R(z) \quad (3)$$

where $n(z)$ and $p(z)$ the electron and hole carrier densities, $j_n(z)$ and $j_p(z)$ the electron and hole current densities, $V(z)$ the electric potential, $\epsilon(z)$ the dielectric constant, and $R(z)$, the recombination-generation rate. The current densities are the sum of a drift and a diffusion component,

$$j_n(z) = -ne\mu_n\frac{dV}{dz} + eD_n\frac{\partial n}{\partial z} \quad (4)$$

$$j_p(z) = -pe\mu_p\frac{dV}{dz} - eD_p\frac{\partial p}{\partial z} \quad (5)$$

where $\mu_n$ and $D_n$ are the electron mobility and diffusivity, and $\mu_p$ and $D_p$ the hole mobility and diffusivity. These quantities are related by the generalized Einstein relations as in Eq. (15) of Ref. 19.

The recombination-generation rate is assumed to be given by $R = R_{SRH} + R_{\text{bim}}$, where $R_{SRH}$ is given by a sum of Shockley-Read-Hall terms:[18, 20]

$$R_{SRH} = \sum_t \frac{[n(z)p(z) - n_i^2]}{\tau_{pt}[n_t + n(z)] + \tau_{nt}[p_t + p(z)]} \quad (6)$$
$$= [n(z)p(z) - n_i^2]r_{SRH}(z),$$

Here $\tau_{nt}$ and $\tau_{pt}$ are the electron and hole recombination lifetime parameters for trap $t$, $n_i$ the intrinsic carrier concentration, and the quantities $n_t$ and $p_t$ are defined as[18]

$$n_t = N_c(T)\exp\left(\frac{E_t - E_c}{k_B T}\right) \quad (7)$$

and

$$p_t = P_v(T)\exp\left(\frac{E_v - E_t}{k_B T}\right) \quad (8)$$

where $N_c(T)$ and $P_v(T)$ are the thermal densities of electrons and holes, respectively, $E_c$ the conduction band edge, $E_v$ the valence band edge, and $E_t$ the trap level energy. From these definitions $n_t$ and $p_t$ can be viewed as the electron and hole concentrations in the conduction and valence bands if the quasi-Fermi levels were at the same energy as the trap level $\epsilon_t$. According to Eqs. (6)-(8), the set of three parameters $(\tau_{nt}, \tau_{pt}, \epsilon_t)$ completely defines the contribution from each trap level to the recombination-generation current. As we will see below, the position of the trap level relative to the band edges has a strong influence not only on the magnitude of this current but also on the overall I-V lineshape.

The bimolecular radiative recombination rate is [20]

$$R_{\text{bim}}(z) = B_{\text{bim}}[n(z)p(z) - n_i^2], \quad (9)$$

where $B_{\text{bim}}$ is the bimolecular recombination coefficient. It is also quite straightforward to include Auger recombination and band to band tunneling (BTBT),[20] but we find these are not nearly as important as SRH recombination.

Our primary goal is to solve for the total current density $j = j_n(z) + j_p(z)$. Generalizations of Eqs. (1)-(5) to higher dimensions are straightforward, but the computational cost would make it extremely difficult to attempt any fit of the I-V characteristics. This is particularly so the case in the forward bias regime, for which serial resistance effects require the solution of the semiconductor equations not once but multiple times in order to numerically solve the nonlinear equation involving the current in a diode with a resistor in series.

From Eqs. (1)-(5) it is apparent that the problem involves three independent functions $\{n(z), p(z), V(z)\}$. For the numerical solution, it is convenient to shift and normalize the potential as

$$v(z) = \frac{eV(z) + F - F_i}{k_B T}, \quad (10)$$

where $F$ is the equilibrium Fermi level and $F_i$ the intrinsic Fermi level, $k_B$ Boltzmann's constant and $T$ the temperature. We can then write the carrier concentrations as

$$\begin{aligned} n(z) &= n_i u_n(z) \exp[v(z)] \\ p(z) &= n_i u_p(z) \exp[-v(z)] \end{aligned} \quad (11)$$

where $n_i$ is the intrinsic carrier concentration. With these definitions, Eqs. (4) and (5) become, using Einstein's relations,

$$j_n(z) = eD_n n_i \exp[v(z)] \frac{du_n(z)}{dz} \quad (12)$$

$$j_p(z) = -eD_p n_i \exp[-v(z)] \frac{du_p(z)}{dz} \quad (13)$$

The function set $\{u_n(z), u_p(z), v(z)\}$ are the so called Slotboom variables and turn out to be convenient for numerical work.(Ref. 21) Their physical meaning becomes apparent if we write

$$\begin{aligned} u_n(z) &= \exp\left[\frac{F_n(z) - F}{k_B T}\right] \\ u_p(z) &= \exp\left[-\frac{F_p(z) - F}{k_B T}\right]. \end{aligned} \quad (14)$$

Then, by comparing with the standard quasi-equilibrium analytical expressions for $n(z)$ and $p(z)$, it is easy to see that $F_n(z)$ and $F_p(z)$ can be identified with the electron and hole quasi-Fermi levels.

Eqs (1)-(3) can be written in terms of $\{u_n(z), u_p(z), v(z)\}$ as

$$-\frac{d}{dz}\left[\epsilon(z)\frac{dv}{dz}\right] = \frac{4\pi e^2 n_i}{k_B T}\left[u_p(z)\exp[-v(z)] \right. \\ - u_n(z)\exp[v(z)] \\ \left. + \frac{N_d(z) - N_a(z)}{n_i}\right] \quad (15)$$

$$\frac{d}{dz}\left\{D_n n_i \exp[v(z)]\frac{du_n(z)}{dz}\right\} = R(z) \quad (16)$$

$$\frac{d}{dz}\left\{D_p n_i \exp[-v(z)]\frac{du_p(z)}{dz}\right\} = R(z) \quad (17)$$

### B. Discretization

Ge diodes with highly-doped quasi-neutral regions are a challenging system for numerical simulations because two very different length scales coexist. One of them is the Debye length $\ell$, relevant in the depletion layer, and the other one is the diffusion length $L_D$, relevant in the quasi-neutral regions. For $n,p \sim 10^{19}$ cm$^{-3}$, typical values of the Debye length are $\ell$=1.5 nm, whereas $L_D >$ 3000 nm. With such a severe mismatch, the need for a non-uniform discretization grid is apparent. The separation between two successive points is denoted as $h_j = z_{j+1} - z_j$. In terms of these quantities, it can be shown that the discrete version of Eqs. (15)-(17) is

$$-\frac{2\epsilon_{j+\frac{1}{2}}}{(h_j + h_{j-1})h_j}v_{j+1} - \frac{2\epsilon_{j-\frac{1}{2}}}{(h_j + h_{j-1})h_{j-1}}v_{j-1} \\ = -\left[\frac{2\epsilon_{j+\frac{1}{2}}}{(h_j + h_{j-1})h_{j+1}} + \frac{2\epsilon_{j-\frac{1}{2}}}{(h_j + h_{j-1})h_{j-1}}\right]v_j \\ + \frac{4\pi e^2}{k_B T}[n_i u_{pj}\exp(-v_j) - n_i u_{nj}\exp(v_j) + N_{dj} - N_{aj}], \quad (18)$$

$$\frac{2n_i D_{n,j+\frac{1}{2}}}{(h_j + h_{j-1})h_j}B(v_j - v_{j+1})\exp(v_j)u_{n,j+1} + \frac{2n_i D_{n,j-\frac{1}{2}}}{(h_j + h_{j-1})h_{j-1}}B(v_j - v_{j-1})\exp(v_j)u_{n,j-1} \\ = \left[\frac{2n_i D_{n,j+\frac{1}{2}}}{(h_j + h_{j-1})h_j}B(v_j - v_{j+1})\exp(v_j) \right. \\ \left. + \frac{2n_i D_{n,j-\frac{1}{2}}}{(h_j + h_{j-1})h_{j-1}}B(v_j - v_{j-1})\exp(v_j)\right]u_{nj} \\ + n_i^2[u_{n,j}u_{p,j} - 1](r_{j,\text{SRH}} + B_{\text{bim}}), \quad (19)$$

and

$$-\frac{2n_i D_{p,j+\frac{1}{2}}}{(h_j+h_{j-1})h_j}B(v_{j+1}-v_j)\exp(-v_j)u_{p,j+1} - \frac{2n_i D_{p,j-\frac{1}{2}}}{(h_j+h_{j-1})h_{j-1}}B(v_{j-1}-v_j)\exp(-v_j)u_{p,j-1}$$

$$= -\left[\frac{2n_i D_{p,j+\frac{1}{2}}}{(h_j+h_{j-1})h_j}B(v_{j+1}-v_j)\exp(-v_j) \right.$$
$$\left. + \frac{2n_i D_{p,j-\frac{1}{2}}}{(h_j+h_{j-1})h_{j-1}}(v_{j-1}-v_j)\exp(-v_j)\right]u_{p,j}$$
$$- n_i^2[u_{n,j}u_{p,j}-1](r_{j,\text{SRH}}+B_{\text{bim}}) \quad (20)$$

where for any function of position, $f_{j+\frac{1}{2}} = f(z_j + \frac{1}{2}h_j)$, and we introduce the Bernoulli generation function

$$B(y) = \frac{y}{\exp(y)-1}. \quad (21)$$

This function appears following a method introduced by Scharfetter and Gummel[22] to avoid unphysical negative carrier concentrations that may arise because the discretization approximation $n_{j+\frac{1}{2}} = \frac{1}{2}(n_{j+1}+n_j)$ is not sufficiently good near the edge of depletion layers, where the carrier concentrations change exponentially.[23] The numerical evaluation of $B(y)$ can also lead to instabilities that we eliminate by expanding the denominator to cubic order for $y < 1.5\times 10^{-3}$.

We will also need to calculate the electric field $-dv/dz$ to account for TAT recombination and the Poole-Frenkel effect. This electric field is required at mesh points, not half-way between points. We then use:[24]

$$-dv/dz|_j = $$
$$-\frac{v_{j+1} - \left(\frac{h_j}{h_{j-1}}\right)v_{j+1} - \left[1-\left(\frac{h_j}{h_{j-1}}\right)^2\right]v_j}{h_j\left(1+\frac{h_j}{h_{j-1}}\right)} \quad (22)$$

Each of the three equations (18)-(20) has the form

$$\alpha_{j+1}\eta_{j+1} + \alpha_{j-1}\eta_{j-1} = \alpha_j\eta_j + G_j(\eta_j) + f_j \quad (23)$$

where $\eta$ is a generic symbol for either $v$, $u_n$ or $u_p$

The three equations can be solved iteratively using a method proposed by Mayergoyz and coworkers.[21, 25] The method consists of solving the equations

$$\Psi(\eta_j^{K+1}) \equiv \alpha_j\eta_j^{K+1} + G_j(\eta_j^{K+1}) - F_j^{K+1} \quad (24)$$
$$= 0$$
$$F_j^{K+1} = \alpha_{j+1}\eta_{j+1}^K + \alpha_{j-1}\eta_{j-1}^K - f_j^K$$

where the superscript $K$ indicates the order of the iteration. The first equation is a nonlinear equation in $\eta_j^{K+1}$, which is solved itself iteratively, using for example the Newton method. This implies two iteration loops, one for $K$ and one to solve the nonlinear equation. However, it makes no sense to spend computer time to solve the nonlinear equation "exactly" if we are only finding an approximate solution for loop $K$. Therefore, we limit ourselves to the first iteration of the inner loop,[26] which leads to

$$\eta_j^{K+1} = \eta_j^K - \frac{\alpha_j\eta_j^K + G_j(\eta_j^K) - F_j}{\alpha_j + G_j'(\eta_j^K)} \quad (25)$$

For the derivatives of $G$, the potential case is straightforward:

$$G'(v_j) = -\frac{4\pi e^2 n_i}{k_B T}\{u_{pj}\exp(-v_j) + u_{nj}\exp(v_j)\} \quad (26)$$

For the current continuity equations, on the other hand, the derivatives are complex because $r_j$ is itself a function of $u_{n,j}$ and $u_{p,j}$. We then use the so called Seidman's modification[27] by writing

$$G_j(u_{n,j}^{K+1}) = n_i^2 u_{p,j}^K u_{n,j}^{K+1}(r_{j,\text{SRH}}^K + B_{j,\text{bim}}^K) \quad (27)$$
$$f_j^K = -n_i^2(r_{j,\text{SRH}}^K + B_{j,\text{bim}}^K)$$

and

$$G_j(u_{p,j}^{K+1}) = -n_i^2 u_{n,j}^K u_{p,j}^{K+1}(r_{j,\text{SRH}}^K + B_{j,\text{bim}}^K) \quad (28)$$
$$f_j^K = -n_i^2(r_{j,\text{SRH}}^K + B_{j,\text{bim}}^K)$$

This gives

$$G'(u_{n,j}^{K+1}) = n_i^2 u_{p,j}^K(r_{j,\text{SRH}}^K + B_{j,\text{bim}}^K) \quad (29)$$
$$G'(u_{p,j}^{K+1}) = -n_i^2 u_{n,j}^K(r_{j,\text{SRH}}^K + B_{j,\text{bim}}^K)$$

The final modification prior to numerical evaluation is to introduce normalized quantities. We define

$$\bar{\epsilon} = \min(\epsilon's) \quad (30)$$
$$\bar{N} = \max(N's)$$
$$D_{\max} = \max(D's)$$

and a length

$$\ell = \sqrt{\frac{\bar{\epsilon}k_B T}{4\pi e^2 \bar{N}}}. \quad (31)$$

This allows us to introduce a dimensionless coordinate $x$ and dimensionless quantities $\epsilon'$, $h'$, and $D'$ such that

$$z = \ell x \quad (32)$$

$$h_j = \ell h'_j$$
$$\epsilon = \bar{\epsilon}\epsilon'$$
$$D = D_{\max}D'$$

Similarly, we can define dimensionless recombination-generation coefficients as

$$r'_j = \frac{n_i \ell^2}{D_{\max}} r_j \tag{33}$$
$$B'_{j,\text{bim}} = \frac{n_i \ell^2}{D_{\max}} B_{j,\text{bim}}$$

A table with the coefficients in Eq. (23) using the normalized quantities just defined is shown in Appendix B.

### C. Boundary conditions and initial guess

To explain the boundary conditions, we assume for concreteness that the $p$-layer is on the left of the diode and extends from $z_- = -W_p - d/2$ to $z_+ = W_n + d/2$, with ohmic contacts at both ends. Accordingly, equilibrium conditions and charge neutrality must prevail at both ends. If no external voltage is applied, equilibrium implies, according to Eq (14), $u_n(z_-) = u_p(z_-) = u_n(z_+) = u_p(z_+) = 1$. Then the charge neutrality condition on the left side is $N_a + n - p = 0$, so that Eq. (11) implies

$$v(z_-) = -\ln\left(\frac{N_a}{2n_i} + \frac{N_a}{2n_i}\sqrt{1 + \frac{4n_i^2}{N_a^2}}\right). \tag{34}$$

Similarly, we obtain

$$v(z_+) = \ln\left(\frac{N_d}{2n_i} + \frac{N_d}{2n_i}\sqrt{1 + \frac{4n_i^2}{N_d^2}}\right) \tag{35}$$

Let us now consider the case when a voltage $v_{\text{app}}$ is applied between the diode terminals. In that case, the boundary conditions for the potential are

$$v(z_-) = \frac{v_{\text{app}}}{2} \tag{36}$$
$$-\ln\left(\frac{N_a}{2n_i} + \frac{N_a}{2n_i}\sqrt{1 + \frac{4n_i^2}{N_a^2}}\right).$$

and

$$v(z_+) = -\frac{v_{\text{app}}}{2} \tag{37}$$
$$+\ln\left(\frac{N_d}{2n_i} + \frac{N_d}{2n_i}\sqrt{1 + \frac{4n_i^2}{N_d^2}}\right)$$

But since the carrier concentrations at the contacts are independent of the voltage at those points, then

$$u_n(z_+) = \exp\left[\frac{v_{\text{app}}}{2}\right] \tag{38}$$

$$u_p(z_+) = \exp\left[-\frac{v_{\text{app}}}{2}\right]$$
$$u_p(z_-) = \exp\left[\frac{v_{\text{app}}}{2}\right]$$

Because the solution method used here is *globally convergent*,[21] any initial guess for the functions $\{u_n(z), u_p(z), v(z)\}$ leads eventually to the correct solution, but the time needed can vary considerably. We find that a reasonably fast solution is found by mimicking the depletion approximation. We assume that the initial potential $v_{\text{initial}}(z)$ is equal to $v(z_-)$ from $z_-$ to $z_{\text{left}} = -d/2 - 3\ell$ and equal to $v(z_+)$ from $z_{\text{right}} = d/2 + 3\ell$ to $z_+$. For $z_{\text{left}} < z < z_{\text{right}}$, we linearly interpolate the potential values between $v(z_-)$ and $v(z_+)$. Similarly, we define a function $v_{\text{app,initial}}(z)$ such that it is equal to $v_{\text{app}}$ for $z < z_{\text{left}}$, equal to $-v_{\text{app}}$ for $z_{\text{right}} < z$, and linearly interpolated between these values for $z_{\text{left}} < z < z_{\text{right}}$. With this definition, we take $u_{n,\text{initial}}(z) = \exp\left[-\frac{v_{\text{app,initial}}(z)}{2}\right]$ and $u_{p,\text{initial}}(z) = \exp\left[\frac{v_{\text{app,initial}}(z)}{2}\right]$.

### D. External circuit

We assume the diode to be embedded in a circuit as depicted in Fig. 1. The externally applied voltage is $V_{\text{ext}}$ and the externally measured current is $I$, which are not the same as the voltage $V_{\text{d}}$ across the diode and the diode current diode $I_d$. The resistors in the figure account for the measuring circuit as well as for the physical characteristics of the diode itself. For example, non-ideal passivation of the side walls may lead not only to additional recombination/generation, as discussed below, but to a non-infinite value of $R_{\text{in}}$. By solving the circuit, we find that

$$V_{\text{d}} = (V_{\text{ext}} - I_d R_s)\left(\frac{R_{\text{in}}}{R_{\text{in}} + R_s}\right). \tag{39}$$

The measured current is

$$I = \left(\frac{R_{\text{in}}}{R_{\text{in}} + R_s}\right)\left(\frac{V_{\text{ext}}}{R_{\text{in}}} + I_d\right) + \frac{V_{\text{ext}}}{R_{\text{out}}} \tag{40}$$

To find the diode current $I_d$ we must solve the nonlinear equation

$$I_d = f\left[(V_{\text{ext}} - I_d R_s)\left(\frac{R_{\text{in}}}{R_{\text{in}} + R_s}\right)\right] \tag{41}$$

For an ideal diode, the solution is provided by the Lambert W-function, but in our case we need a numerical solution in which, for every iteration of the root-finding method, we must solve the semiconductor differential equations. The serial resistance can almost never be neglected, and in some cases it may be increased intentionally to measure properties such as responsivity. Thus, accounting for its effect on the I-V curves is by far the most time-consuming aspect of the calculation, but it is essential to match experimental curves.

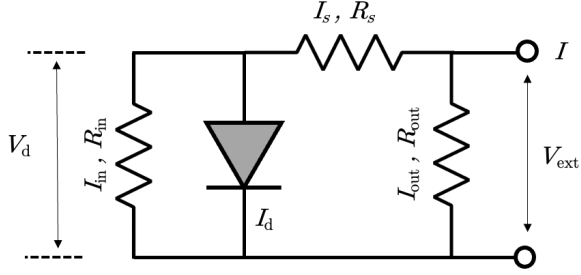

**Fig. 1** Schematic diagram of the diode circuit leading to Eq. (40).

We incorporate the differential equation solver into a root finding algorithm using Brent's method,[28] as implemented in the commercial package Igor Pro from Wavemetrics Inc. Even with these complications, the time needed to compute a full I-V curve can be as short as a minute in a fast personal computer.

**E. Recombination model details**

*1. Trap-assisted-tunneling and Poole-Frenkel effect*

As indicated above, we assume that all defect-related recombination-generation processes can be described in terms of SRH terms. To account for trap-assisted tunneling, we follow the work of Hurkx[29] and rewrite each of the summation terms in Eq. (6) as

$$\Delta E_n(z) = \begin{cases} E_g - E_t & \text{for } E_t \geq E_g - e[V(z_{\text{right}}) - V(z)] \\ e[V(z_{\text{right}}) - V(z)] & \text{for } E_t < E_g - e[V(z_{\text{right}}) - V(z)] \end{cases} \quad (45)$$

and

$$\Delta E_p(z) = \begin{cases} e[V(z) - V(z_{\text{left}})] & \text{for } E_t \geq e[V(z) - V(z_{\text{left}})] \\ E_t & \text{for } E_t < e[V(z) - V(z_{\text{left}})] \end{cases} \quad (46)$$

These barriers may be lowered by the Poole-Frenkel effect, caused by the presence of an electric field of magnitude $|\mathbf{E}|$. We can write this as

$$\Delta E_{PF} = \beta_{PF}|\mathbf{E}| \quad (47)$$

For a Coulombic trap, one can show that $\beta_{PF} = \sqrt{4e^3/\epsilon}$. The barriers in Eqs. (45) and (46) are modified to new values, written as $\Delta E'_{n,p}$, which for a donor trap are given as

$$\Delta E'_n = \Delta E_n - \Delta E_{PF} \quad (48)$$
$$\Delta E'_p = \Delta E_p.$$

Similarly, for an acceptor trap we obtain

$$\Delta E'_n = \Delta E_n \quad (49)$$

$$R_t(z) = \quad (42)$$
$$= \frac{[n(z)p(z) - n_i^2]}{\frac{\tau_{pt}}{1+\Gamma_{pt}}[n_t + n(z)] + \frac{\tau_{nt}}{1+\Gamma_{nt}}[p_t + p(z)]}$$

Here the recombination enhancement functions $\Gamma_{nt}(z)$ and $\Gamma_{pt}(z)$ account for trap assisted tunneling (TAT) and the Poole-Frenkel effect, as described below.

Trap-assisted-tunneling refers to carrier tunneling through the approximately triangular potential barrier in the presence of the electric field. We model the TAT contribution using the expressions derived by Hurkx:[29]

$$\Gamma_{nt} = \frac{\Delta E_n}{k_B T} \int_0^1 \exp\left(\frac{\Delta E_{nt}}{k_B T}u - K_{nt}u^{3/2}\right) du$$
$$\Gamma_{pt} = \frac{\Delta E_p}{k_B T} \int_0^1 \exp\left(\frac{\Delta E_{pt}}{k_B T}u - K_{pt}u^{3/2}\right) du \quad (43)$$

with

$$K_{nt} = \frac{4}{3}\frac{\sqrt{2m_\parallel(\Delta E_{nt})^3}}{e\hbar|\mathbf{E}|}$$
$$K_{pt} = \frac{4}{3}\frac{\sqrt{2m_\parallel(\Delta E_{pt})^3}}{e\hbar|\mathbf{E}|} \quad (44)$$

Where $|\mathbf{E}|$ is the magnitude of the electric field, $m_\parallel$ is the effective mass in the direction of transport, and $\hbar$ the reduced Planck constant. The quantities $\Delta E_n$ and $\Delta E_p$ are the tunneling barriers and they are defined as (if we take $E_v = 0$, $E_c = E_g$):

$$\Delta E'_p = \Delta E_p - \Delta E_{PF}.$$

*2. Perimetral recombination-generation*

The diode model presented so far predicts that diodes that are identical except for their areas should have identical *j*-V curves. This is hardly ever observed in real life, the main reason being the presence of side wall recombination-generation that is not proportional to the area but to the diodes' perimeter. Good passivation can mitigate the effect, but not to the extent that perimetral recombination can be ignored if we seek a quantitative reproduction of the experimental I-V curves.

Unfortunately, incorporating side wall recombination in a rigorous way requires a three-dimensional diode model. This, as indicated above, is incompatible with our stated goal of developing a fast, quantitative diode model that can be used to fit experimental curves. Therefore, we account for perimetral recombination-generation by developing a one-dimensional effective model. Starting from the fully three-dimensional semiconductor equations, we define average quantities by integrating over the diode area. These average quantities are only functions of the coordinate $z$ and satisfy the one-dimensional equations (1)-(5). If we assume that perimetral defects are located within a distance $\delta r$ of the diode surface, the averaging process for the SRH term corresponding to perimetral recombination gives

$$R_{\text{per}}(z) = \left(\frac{4}{D}\right) \times \frac{n(z)p(z) - n_i^2}{1/v_{n,\text{per}}[p_{\text{per}} + p(z)] + 1/v_{p,\text{per}}[n_{\text{per}} + n(z)]} \quad (50)$$

Here $D$ is the device's diameter and we introduce perimetral recombination velocities $v_{n,\text{per}} = \delta r/\tau_{n,\text{per}}$, $v_{p,\text{per}} = \delta r/\tau_{p,\text{per}}$.

### 3. Recombination parameter choice

Our model contains one SRH term for each trap present. Unfortunately, a full independent identification of all such traps and their microscopic structure is not available. Therefore, our philosophy will be to use a model containing the smallest possible set of traps that is compatible with all available diode experimental data. We find that that this requires four trap levels. One trap is assumed to be the same one that appears in bulk Ge and represents the dominant recombination mechanism in Ge solar cells on Ge substrates. A second trap is associated with the above perimetral recombination. The corresponding trap level is assumed to be located in middle of the band gap $E_g$, at an energy $E_g/2$ above the valence band edge, simply because these is the trap energy that maximizes SRH recombination. In addition to these traps, we find that for Ge-on-Si we need to include two additional dislocation-related traps. One of these traps is assumed to be at the same mid-gap location as the above traps, the other one at an energy $E_g/4$ above the valence band edge. There are several defect studies in Ge-on-Si as well as GeSn-on-Si which provide evidence for shallow defects above the valence band edge at energies close to $E_g/4$,[11,13,30-32] and we will show below that the presence of one such level is important for fitting the shape of the forward-bias I-V curves.

For each of the above traps, we need the electron- and hole recombination lifetimes $\{\tau_{nt}, \tau_{pt}\}$ (or the surface recombination velocities $\{v_{n,\text{per}}, v_{p,\text{per}}\}$ in the case of perimetral recombination) to compute the corresponding $R_{SRH}$ term. For bulk Ge recombination there are some lifetime data in the literature, and we use those values, as discussed in Appendix A. Much less is known about dislocation-related recombination. We assume that electron- and hole-recombination lifetimes are the same and inversely proportional to the dislocation density $\rho_D$:

$$\tau_{n,\text{mid}} = \tau_{p,\text{mid}} = \frac{1.2}{\rho_D}\left(\frac{\text{s}}{\text{cm}^2}\right) \quad (51)$$

A discussion of this choice is given in Appendix A. We also assume that the shallow level recombination parameters are

$$\tau_{n,\text{shallow}} = \tau_{p,\text{shallow}} = \frac{\tau_{n,\text{mid}}}{r_{\text{md}}} = \frac{\tau_{p,\text{mid}}}{r_{\text{md}}} \quad (52)$$

Where $r_{\text{md}}$ is a number that we expect to be approximately constant among different Ge-on-Si samples if our defect model is physically valid. In fact, we find that the choice $r_{\text{md}} = 75$ leads to good agreement with experiment in most cases.

Using the final assumption $v_{n,\text{per}} = v_{p,\text{per}}$, we then end up with essentially a two-parameter recombination model in which the main adjustable parameters are the dislocation density and the surface recombination velocity. It is apparent that some of the assumptions made in our trap models are crude and ad hoc, but more realistic accounts require very detailed defect studies that are simply not available at this time.

## III. MODEL PREDICTIONS

Before attempting to fit real experimental data, we show some model predictions that illustrate the perils of "naïve" studies based on analytical approximations, as described in the introduction, and show how the different model parameters affect the predicted I-V curves.

### A. Electric field effects

Figure 2 shows a few calculated electric fields for a typical *pin* structure. The accurate calculation of these fields is important, among other considerations, because they modify recombination via the TAT and Poole-Frenkel effects. We consider the cases in which the *i*-layer is truly intrinsic and those in which there is some residual carrier concentration. Even in the purely intrinsic case, the calculated electric field is significantly different from the depletion approximation analytical prediction near the *n/i* and *p/i* interfaces, as is well known.[18] If residual doping is present, the electric field is drastically changed, and for this situation there are no simple analytical approximations that can be used. These simulations underscore the need for a fully numerical solution of the semiconductor equations.

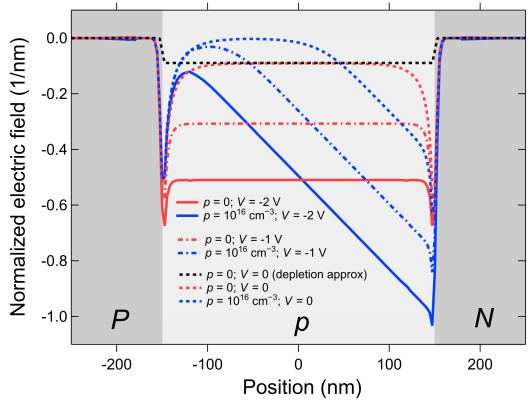

**Fig. 2** Calculated electric fields in a *pin* diode ($P = N = 10^{19}$ cm$^{-3}$) at three different applied voltages. Solid lines are for $V_{\text{app}} = -2$ V, dash-dotted lines for $V_{\text{app}} = -1$ V, and dotted lines for $V_{\text{app}} = 0$. At each voltage, the red line corresponds to a perfectly intrinsic layer and the blue line to an intrinsic layer with a residual doping concentration $p = 10^{16}$ cm$^{-3}$. The black dotted line is the depletion approximation result for $V_{\text{app}} = 0$.

### B. Voltage dependence of reverse bias current

An important aspect of the experimental I-V curves in Ge-on-Si diodes is the relatively strong voltage dependence of the reverse-bias current. Fig. 3 shows how the TAT and Poole-Frenkel mechanisms affect the slope of the reverse-bias curve. When residual doping is present, Fig. 2 shows that in the nominally intrinsic layer there is a region near the *i/n* interface (for hole residual carriers) with higher electric fields compared to the case of no residual doping, and a region with very low electric fields that decreases as the reverse bias voltage increases. This causes the strongest voltage dependence of the reverse-bias curve, as shown by the red dash-dotted curve in Fig. 3. Because of this dependence, the amount of residual doping is a sensitive and critical parameter to fit experimental I-V curves.

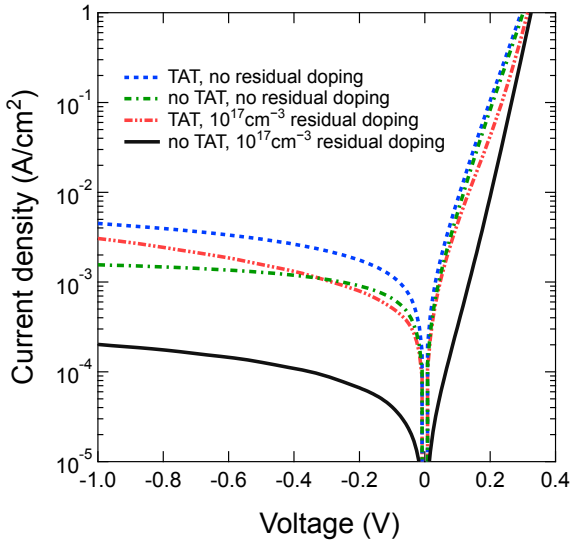

**Fig. 3** Calculated *j*-V curves for a Ge *PiN* diode ($P = N = 10^{19}$ cm$^{-3}$). If there is no residual doping in the intrinsic layer, the dash-dotted green and dotted blue lines show the enhancement of the reverse-bias current due to TAT processes. In the presence of residual doping in the intrinsic layer ($p = 10^{17}$ cm$^{-3}$) there is a significant reduction of the reverse-bias current if TAT is absent (solid black line), but including TAT enhances the reverse-bias current and increases its slope.

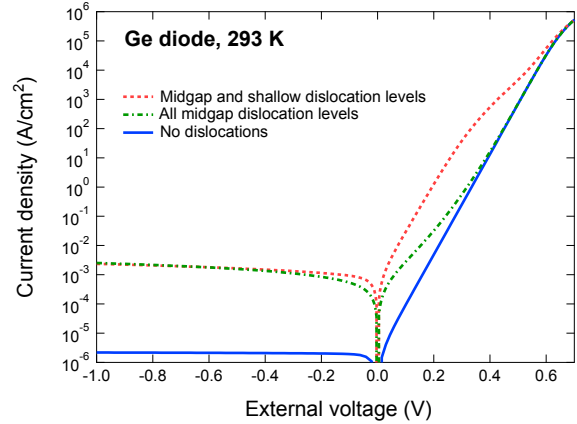

**Fig. 4** Calculated *j*-V curves for a Ge *PiN* diode ($P = N = 10^{19}$ cm$^{-3}$). The solid blue line is the prediction for the case where there are no dislocation defects. The green dash-dotted curve is the prediction if all dislocation-related traps are at midgap, while the red dotted curve assumes the presence of shallow dislocation traps. The dislocation densities were adjusted to match the reverse-bias curves in order to highlight the differences under forward bias.

### C. Trap-energy dependence of the forward bias current

Figure 4 shows calculated *j*-V curves assuming different energies for the trap levels associated with dislocations. The dislocation densities for the two simulations have been adjusted to match the reverse bias curves. Since the curves are very similar in this range, it is possible to fit experimental data with very different ratios of mid-gap to shallow traps. On the other hand, in the forward bias regime the curves look very different, and therefore the forward bias regime is sensitive to the choice of trap level, as anticipated above.

### D. Temperature dependences

Figure 5 shows Arrhenius plots from *j*-V curves calculated with our model for the case of no defects, when the reverse bias current is purely diffusive. In this case, an elementary analysis shows that the activation energy $E_a$ is the material's band gap $E_g = 0.66$ eV. However, a fit of the Arrhenius plot gives a much higher activation energy $E_a = 0.81$ eV. We recover an activation energy very close to the band gap value if we "freeze" the band structure parameters, mobilities, and lifetimes to their values at 20 °C. Similarly, we show in Fig 6 the same calculation but for a case when the reverse-bias current is dominated by mid-gap defects, in which case the "expected" activation energy is $E_a = E_g/2$. Again, we obtain a result very close to this value ($E_a = 0.32$ eV) when we "freeze" the electronic and transport parameters at 20 °C, but when the full temperature dependences are included, the predicted activation energy is $E_a = 0.42$ eV, a considerable higher value. These results suggest that simple Arrhenius plots of experimental data are a poor way to determine the relevant trap energies. This has already been recognized in the literature, although many authors rely on such methods for their analysis. Modified methodologies have been proposed to study the effect of traps on the diode current.[33] Our model suggests a complementary and potentially more powerful approach, which is to compare the predicted and observed temperature dependence of the reverse bias current and adjust the trap energies until a good match between theory and experiment is obtained.

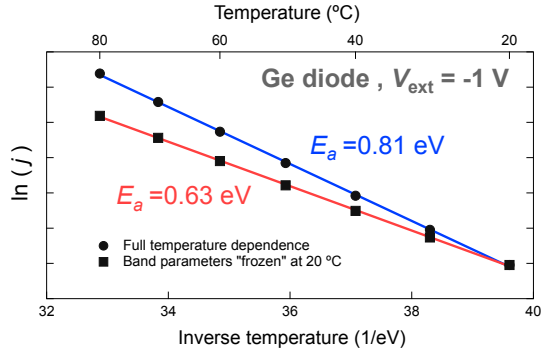

**Fig. 5** Arrhenius plots of the predicted reverse-bias current at -1V for a Ge *pin* diode containing no defects. Circles correspond to the full calculation, which gives an activation energy $E_a = 0.81$ eV. Squares correspond to the same calculation but keeping the electronic structure parameters artificially held at 20 °C. The activation energy is then reduced to $E_a = 0.63$ eV, which is very close to the band gap $E_g = 0.66$ eV.

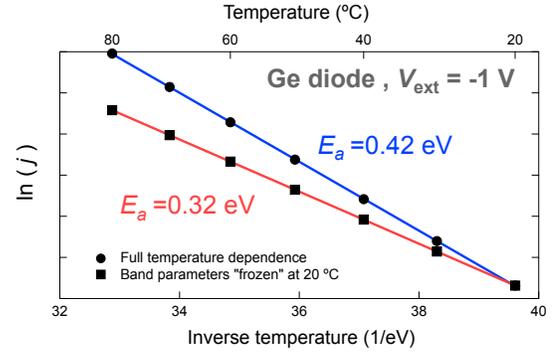

**Fig. 6** Arrhenius plots of the predicted reverse-bias current at -1V for a Ge *pin* diode for which the current is dominated by midgap defects. Circles correspond to the full calculation, which gives an activation energy $E_a = 0.42$ eV. Squares correspond to the same calculation but keeping the electronic structure parameters artificially held at 20 °C. The activation energy is then reduced to $E_a = 0.32$ eV, which is very close to $E_g/2$.

## IV. EXPERIMENTAL DATA FITS

While there are numerous publications reporting I-V curves for Ge-on-Si diodes, many of those articles fail to report these curves in sufficient detail and to provide the essential sample parameters that are needed for a simulation with our model. These parameters include layer thicknesses and doping levels. Strain also modifies the band structure and affects the I-V curves. Our model accounts for some of those effects (See Appendix A) Similarly, I-V curves from diodes of different sizes are critical to separate the perimetral from bulk contributions to the defect currents. Furthermore, as we will demonstrate below, measurements as a function of temperature are important to accurately fit the parallel resistances in Fig. 1. We show next a few examples from Ge-on-Si samples grown following different methods.

### A. Two-step growth method

Perhaps the most popular approach to grow low-defect Ge on Si consists in a two-step approach in which an initiation Ge layer is deposited at relatively low temperatures around 400 °C followed by a growth temperature ramp up to about 670 °C.[34] Complete Ge homostructure *pin* diodes fabricated using this method were characterized by Colace *et al*.[35] Figure 7 shows the *j*-V curve for $D = 80$ μm device. Since the current density is about the same for all device sizes, we set the surface recombination velocities equal to zero for our sample modeling. The solid line in Fig. 7(a) shows our fit using a dislocation density $\rho_D = 8\times10^6$ cm$^{-2}$ and $r_D = 100$. We assume no residual doping in the intrinsic layer and a serial resistance of 400 Ω. We see that the fit is in excellent agreement with the data. The dislocation density is within the range reported for this type of growth, and the fact that a good fit is obtained with no residual doping is consistent with SIMS measurements that show dopant densities below $2\times10^{14}$ cm$^{-3}$.

dominated regime. In fact, our model predicts in this case an activation energy $E_a$ = 0.52 eV, entirely due to the defect levels. This value is not in excellent agreement with the experimental measurement, but it is clearly higher than $E_g/2$. Another discrepancy between model and experiment is that Balbi et al. find a strong voltage dependence of the activation energy, with a significant decrease from $V_{ext}$ = -1 to $V_{ext}$ = -3 , for which the activation energy is found to be $E_a$ = 0.32 eV. We find a much weaker voltage dependence from $E_a$ = 0.52 eV at $V_{ext}$ = -1 to $E_a$ = 0.50 eV at $V_{ext}$ = -3 .

### B. Germanium-on-insulator

Son and coworkers[14] have published a detailed study of *pin* photodetectors grown on Ge-on-insulator. They consider samples with high and low dislocation densities, the latter obtained by annealing. They also provide data on size and temperature dependence, which allow for a more complete deployment of a simulation model.

Fig. 8 shows experimental data and model fits for samples with measured high and low threading dislocation densities $\rho_D$ =5.2×10$^8$ cm$^{-2}$ and $\rho_D$ =3.2×10$^6$ cm$^{-2}$, respectively, and different sizes. Our model calculations reproduce the experimental I-V curves using $\rho_D$ =1.1×10$^9$ cm$^{-2}$ and $\rho_D$ =1.6×10$^6$ cm$^{-2}$, within a factor of 2 in both cases. Using perimetral recombination velocities $v_p = v_n$ = 3.4×10$^4$ cm/s we are able to match the size dependence of the reverse bias current in the low-dislocation samples and reproduce the lack of such dependence in the highly dislocated samples. On the forward bias region, we use series resistances of 12 Ω and 15 Ω, respectively, for the $D$ = 250 μm and $D$ = 60 μm

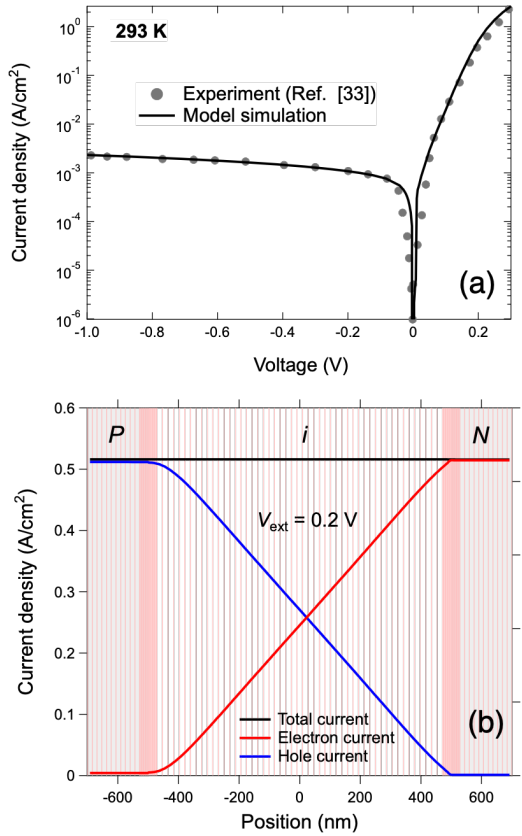

**Fig. 7** (a) Experimental *j*-V curve for a $D$ = 80 μm *pin* diode from Colace et al. (Ref. [33]). The solid line is a simulation assuming no perimetral recombination, a dislocation density $\rho_D$ =8×10$^6$ cm$^{-2}$, a mid-shallow ratio $r_D$=100, and a serial resistance $R_s$ =400 Ω.
(b) Details of the numerical simulation to compute the diode current. The vertical lines show the position of the 86 points in the selected integration mesh. The region with a high-density of points corresponds to the transitions from *P* to *i* and from *i* to *N*. The lines show the electron, hole, and total currents in the structure for an applied voltage of 0.2 eV. This illustrates the degree of convergence of the calculation.

Balbi et al. (Ref. 36) studied the temperature dependence of the I-V curves in samples similar to those described in Ref. 35. For $V_{ext}$ = -1 V, their experimental activation energy is $E_a$ = 0.64 eV, which they attributed to a significant role of diffusion. This is hard to understand because the observed reverse bias currents are three orders of magnitude higher than those our model predicts for a diode without dislocation defects. On the other hand, our simulations, as shown in section III.D, indicate that the temperature dependence of the band structure leads to effective activation energies that are higher than those expected from simple models, so that the value $E_a$ = 0.64 eV does not necessarily imply a diffusion-devices with low levels of dislocations. For the highly dislocated diodes we find that the best fit is obtained with a serial resistance of 45 Ω for the $D = 250$ μm diode, but as high as 2 kΩ for the $D = 60$ μm. We see that for this sample the fit at the highest forward bias is not nearly as good, suggesting some contact anomaly in this particular diode.

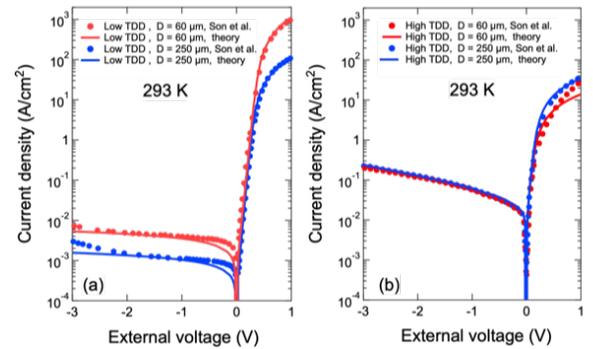

**Fig. 8** Experimental *j*-V curves for *pin* diodes of different diameters $D$ and dislocation densities $\rho_D$ from Son et al. (Ref. 14). Dots are experimental data and solid lines are simulations (a) Diodes with low dislocation levels, fit with $\rho_D$ = 3.2×10$^6$ cm$^{-2}$ and $v_p = v_n$ = 3.4×10$^4$ cm/s ; (b) Diodes with high dislocation levels, fit with $\rho_D$ = 1.1×10$^9$ cm$^{-2}$, $v_p = v_n$ = 3.4×10$^4$ cm/s, and a residual intrinsic layer doping $p$ = 1.5×10$^{17}$ cm$^{-3}$.

The most important qualitative difference between the high-and low dislocation diodes is the observation of a much stronger voltage dependence of the reverse bias current in the former. We reproduce this behavior by assuming some residual doping in the intrinsic layer of the highly dislocated materials; the best fit being obtained using $p = 1.5\times10^{17}$ cm$^{-3}$. On the other hand, in the low-dislocation diodes there is a significant increase in the current for reverse biases beyond 2 eV. This behavior is not reproduced by our model at this point and may be due to BTBT effects.

As to the temperature dependence, an Arrhenius plot of the experimental reverse bias current at -1 V gives for the high-dislocation samples an activation energy $E_a = 0.35$ eV, while our simulations give $E_a = 0.38$ eV, in very good agreement with experiment. On the other hand, for the low-dislocation diodes our model predicts slightly lower activation energies of $E_a = 0.34$-$0.35$ eV, but the Arrhenius plots of the experimental data have a significant non linear component.

### C. Low-temperature Ge precursors

Kouvetakis and coworkers have introduced several low-temperature advanced precursors for the growth of Ge-on-Si that eliminate the need for a two-step process and are compatible with several doping precursors, making it possible to grow in situ complete diode structures with box-like doping profiles.[17, 37-41] The availability of these materials has enabled significant advances in our basic understanding of doped Ge materials, including the study of incomplete ionization,[42] the physics of the doping dependence of the lattice parameter,[43] the observation of Pauli-blocking singularities in the dielectric function of highly doped *n*-Ge,[44, 45] and the advancement of metrology efforts for the determination of carrier concentrations using Hall and optical measurements.[46]

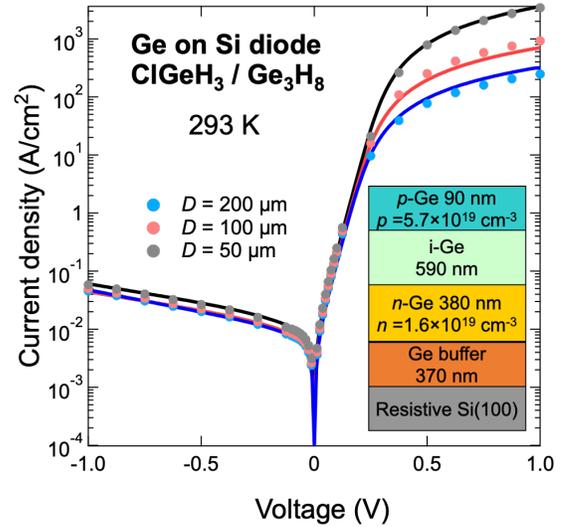

**Fig. 10** Experimental $j$-V curve for a Ge *pin* diode on Si grown using ClGeH$_3$ and Ge$_3$H$_8$ and in situ doping. The solid lines are the experimental curves and the dots correspond to the theoretical simulation. The schematic diagram on the right shows the sample design.

Figure 9 shows selected I-V curves for a device grown by combining Gas Source Molecular Epitaxy (GSME) and Ultrahigh Vacuum Chemical Vapor Deposition (UHV-CVD). Both capabilities are described elsewhere.[47, 48] First a Ge buffer layer, the *n*-type layer, and the intrinsic layer are grown by GSME on a Si (100) substrate using the Ge$_4$H$_{10}$ and P(SiH$_3$)$_3$ precursors at a temperature of 370 °C. The use and advantages of Ge$_4$H$_{10}$ for the growth of Ge were first discussed by Xu *et al.*[47] The sample is then transferred to the UHV-CVD system, where the final *p*-layer is grown using the Ge$_3$H$_8$ and B$_2$H$_6$ precursors. A very good simultaneous fit of the *j*-V curves is obtained using $\rho_D = 10^8$ cm$^{-2}$, $r_{md} = 75$, an intrinsic layer unintentional doping of $p = 9.5\times10^{17}$ cm$^{-3}$, $v_p = v_n = 10^6$ cm/s, and $R_s = 3\Omega$, $R_{in} = \infty$, and $R_{out} = \infty$.

Figure 10 shows the I-V characteristics of a diode that is structurally similar to the previous diode except that the Ge$_4$H$_{10}$ precursor has been replaced with chlorogermane ClGeH$_3$, a new Ge-source recently introduced by our group.[41] A good simultaneous fit of the *j*-V curves is obtained using $\rho_D = 5\times10^8$ cm$^{-2}$, $r_{\mathrm{md}} = 75$, an intrinsic layer unintentional doping of $p = 4\times10^{17}$ cm$^{-3}$, $v_p = v_n = 4\times10^5$ cm/s, and $R_s = 8.6$ Ω, $R_{\mathrm{in}} = \infty$, and $R_{\mathrm{out}} = \infty$.

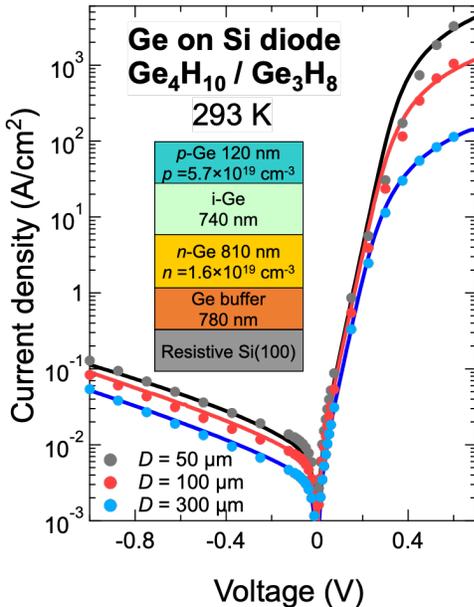

**Fig. 9** Experimental $j$-V curve for a Ge *pin* diode on Si grown using heavy polygermanes and in situ doping. The solid lines are the experimental curves and the dots correspond to the theoretical simulation. The inset shows the schematics of the sample.

While good fits are obtained assuming infinite parallel resistances—attributing all the observed diameter dependence to perimetral recombination, as we did in Figs. 9 and 10—it is difficult as a practical matter to rule out the presence of a finite parallel resistance that is proportional to $1/D$, as discussed in Sect. IIE2. Furthermore, even if one accepts that the diameter dependence is fully accounted for by perimetral recombination, the presence of a finite, diameter independent parallel resistance would affect the slope of the reverse bias curve in a way that is hard to disentangle from the effect of residual intrinsic layer doping combined with TAT processes, as discussed in Section IIIB. For an unambiguous identification of the possible parallel resistance contributions, a study of the temperature dependence of the I-V curve is very useful, as we show next.

Fig. 11 presents the I-V characteristics of a diode grown entirely in the GSME system using Ge$_4$H$_{10}$ as the Ge-precursor. In this case, the $n$-type layer, doped with P(SiH$_3$)$_3$, was grown directly on Si without buffer layers, and the dopant for the top $p$-type layer was the Ga hydride [D$_2$GaN(CH$_3$)$_2$]$_2$ diluted in H$_2$, as described in Ref. 49.

The diode in Fig. 11 displays lower values of the reverse-bias current at room temperature compared to those in Fig. 9 and 10. This implies that the resistances $R_{\mathrm{in}}$ and $R_{\mathrm{out}}$ in Fig. 1 must be both greater than 50 MΩ for the measured current not be affected by their values. For smaller values, the current through the resistors must be accounted for by the model. But as the temperature increases, the diode current increases exponentially and eventually becomes completely dominant over the current through the parallel resistors. Therefore, the diode parameters can be obtained by fits at the higher temperatures, and the parallel resistance values can then be adjusted to improve the fit of the low-temperature I-V curves. This procedure has led to the excellent agreement shown in Fig. 11(a), using $\rho_D = 7.3\times10^7$ cm$^{-2}$, $r_{\mathrm{md}} = 75$, an intrinsic layer unintentional doping of $p = 5\times10^{16}$ cm$^{-3}$, $R_s = 15$ Ω, $R_{\mathrm{in}} = 4$ MΩ, and $R_{\mathrm{out}} = \infty$. An equally good fit is obtained by setting $R_{\mathrm{in}} = \infty$ and adjusting $R_{\mathrm{out}}$. The agreement can be made even better by minor tweaks in the parameters for each of the curves, or by allowing for small deviations in the target temperature within the error of the measurement. Figure 11(b) shows all measured (gray squares) and calculated (red circles) current densities at -1V as a function of temperature. We see that the agreement between model and experiment is quite remarkable. The blue triangles in the figure show a calculation with the same parameters except that we take $R_{\mathrm{in}} = \infty$. A comparison of the red-circle and blue triangle calculations clearly illustrates the aforementioned impact of a non-infinite parallel resistance: at the lowest temperatures there is a clear deviation between triangles and circles, but at higher temperatures the two predictions are virtually identical. These results suggests that some reported deviations from the linear behavior expected in an Arrhenius plot may be simply due to the presence of a non-infinite parallel resistance that has not be properly accounted for. On the other hand, even the blue triangles show some deviation from a purely linear behavior, clearly reflecting the influence of the non-exponential temperature dependence of several material parameters, as discussed above. Nevertheless, we have fit the prediction with a straight line and find an activation energy $E_a = 0.60$ eV. This suggests a diffusion-dominated current, but, as indicated in Sect. IIID, the activation energies obtained from such an exercise cannot be easily interpreted in those terms. In fact, the model calculation shows that at all temperatures, the reverse-bias current is dominated by the defect contribution.

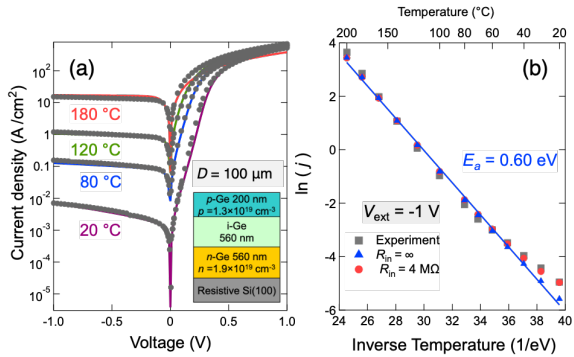

**Fig. 11** (a) Selected experimental $j$-V curves for a Ge $pin$ diode on Si grown entirely in a GSME reactor. The lines are the experimental data and the dots the model predictions using exactly the same parameters $\rho_D = 7.3\times10^7$ cm$^{-2}$, $r_{\mathrm{md}} = 75$, an intrinsic layer unintentional doping of $p = 5\times10^{16}$ cm$^{-3}$, $R_s = 15$ Ω, $R_{\mathrm{in}} = 4$ MΩ, and $R_{\mathrm{out}} = \infty$. (b) Grey squares: Arrhenius plot of the experimental reverse bias current at all temperatures. Red dots: model predictions with the above parameters. Blue triangles: model predictions with the same parameters except that we set $R_{\mathrm{in}} = \infty$. The solid blue line is a linear fit that gives an activation energy $E_a = 0.60$ eV. This is close to the material's band gap, but the reverse-bias current is completely dominated by the defect contribution.

## V. CONCLUSIONS

We have a presented a practical yet realistic model for the I-V characteristics of Ge-on-Si $pin$ diodes. The model can reproduce I-V curves in the literature from a variety of sources by adjusting a very limited number of parameters, and should therefore be useful for systematic studies aimed at improving the performance of such diodes. Extensions to GeSn and SiGeSn materials are straightforward and only require new sets of material parameters instead of those presented in Appendix A, which apply to Ge. Heterostructure diodes can also be included with minor modifications. We will also show in a forthcoming manuscript how the model can be combined with rigorous optical calculations to predict the optical responsivity from Ge, GeSn and SiGeSn *pin* diodes without the need of adjustable parameters such as collection efficiencies.

While the agreement between calculated and experimental I-V curves is excellent, mixed results are obtained for the temperature dependence of these curves. There are some obvious reasons why the predicted temperature dependences may disagree with experiment. For example, we assume the trap levels to track the temperature dependence of the band gap. Different results are obtained in the opposite limit in which the trap level is assumed to lie at a fixed energy relative to one of the band edges. Furthermore, we take the recombination lifetimes for the dislocation-related traps to be temperature-independent for lack of any direct measurement, although there is ample evidence that other recombination lifetimes in the system have their own temperature dependence. From this perspective, a failure to reproduce the temperature dependence of the I-V curves would not necessarily imply that there is a fundamental flaw in our model, but simply reflect our incomplete knowledge of the fundamental physics associated with recombination-generation processes. On the other hand, the excellent agreement between modeled and experimental temperature dependence for some of the modeled diodes (for example Fig. 11) is hard to explain if we are omitting some important contribution to the temperature dependence. In fact, it suggests that better control of the experimental conditions may be necessary to carry out these critical measurements, and that fits that do not account for the parallel resistances may not converge well or lead to erroneous conclusions.

## ACKNOWLEDGMENT

This work was supported by the US National Science Foundation under grant DMR-2119583 and by the Air Force Office of Scientific Research under grant FA9550-23-1-0285. The codes developed under this support run in the Igor Pro platform (Wavemetrics Inc.) and are available upon request for non-commercial purposes.

## APPENDIX A: MODEL PARAMETERS

### D. Lattice parameter

The temperature dependence of the lattice parameters of both Si and Ge are needed to compute thermal stresses on epitaxial films and to evaluate the temperature dependence of the momentum matrix elements, as discussed below. Experimentally, the temperature-dependent lattice parameter can be computed from thermal expansivity measurements. In Ref. 50 we present a physically motivated model that reproduces these experimental data for Si, Ge, and α-Sn.

However, since the expressions derived require numerical integrations to obtain the desired lattice parameter, we take a more practical route here. In the case of Si, we fit the experimental thermal expansion data from Ibach,[51] which extend from 10 K to 800 K, with an eight-order polynomial. This leads to

$$a_{\text{Si}}(T) = a_{0,\text{Si}} \exp\left(\sum_{m=1}^{8} \frac{1}{m}\alpha_m T^m\right) \quad (53)$$

with $\alpha_1$ =3.77292×10$^{-7}$ K$^{-1}$, $\alpha_2$ =-2.42169×10$^{-8}$ K$^{-2}$, $\alpha_3$ =1.84159×10$^{-10}$ K$^{-3}$, $\alpha_4$ =1.75244×10$^{-13}$ K$^{-4}$, $\alpha_5$ =-3.33374×10$^{-15}$ K$^{-5}$, $\alpha_6$ =8.74587×10$^{-18}$ K$^{-6}$, $\alpha_7$ =-9.35695×10$^{-21}$ K$^{-7}$, $\alpha_8$ =3.66327×10$^{-24}$ K$^{-8}$. We also chose $a_0$ =5.4297 Å to match our measurements of bulk Si wafers at room temperature.

For Ge, the available thermal expansion data only reaches 300 K, but ab initio calculations that extend beyond this temperature and agree with the experimental data are available.[52] We have then first adjusted an expression similar to Eq. (53) to the experimental data, and then fit the resulting $a(T)$ curve with an empirical expression of the form

$$a_{\text{Ge}}(T) = a_{0,\text{Ge}} + A\left[1 + \frac{2}{\exp\left(\frac{\Theta}{T}\right) - 1}\right] \quad (54)$$

The fit parameters are $A$ =6.5741×10$^{-3}$ K, and $\Theta$ =355.14 K. The zero-temperature value was chosen as $a_{0,\text{Ge}}$ =5.645 Å to match literature values at room temperature. With these parameters, we obtain a good agreement with the theoretical predictions above 300 K.

### E. Electronic and optical parameters

#### 4. Band gaps

The most important parameter is the fundamental indirect gap, for which we use

$$E_g(T) = E_g(0) - \frac{\alpha T^2}{\beta + T} \quad (55)$$

with $E_g(0)$ =0.7440 eV, $\alpha$ = 4.956×10$^{-4}$ eV/K, and $\beta$ = 216.8 K, as discussed in Ref. 19.

For the direct band gap we use Eq. (27) in Ref. 53, with parameters given in that paper.

#### 5. Effective masses

The valley with minimum at the $L$-point of the Brillouin zone where electron carriers reside in Ge is characterized by a longitudinal mass $m_l$ (corresponding to the <111> directions) and a transverse mass $m_t$. The longitudinal mass is taken as $m_l$= 1.58$m_0$, where $m_0$ is the free electron mass. It is assumed to be independent of temperature.[54] Expressions for the transverse mass as a function of temperature are given in Appendix B of Ref. 45. At 20 °C, the corresponding value is $m_t = 0.0786 m_0$. For calculations involving TAT tunneling, we need effective masses in the direction of the current, which we indicate with a subscript "$c$". For our diodes, the relevant direction is [001]. It is then easy to show that the corresponding electron mass is

$$\frac{1}{m_{\parallel,c}} = \frac{1}{3}\left(\frac{2}{m_t} + \frac{1}{m_l}\right) \quad (56)$$

At the Γ-minimum of the conduction band the dispersion is assumed to be isotropic, so there is no distinction between directions, and the effective mass is given by

$$\frac{m_0}{m_c} = 1 + \frac{2P^2}{3m_0}\left(\frac{2}{E_0} + \frac{1}{E_0 + \Delta_0}\right) \quad (57)$$

where $E_0$ is the direct band gap and $\Delta_0$ the valence band spin-orbit splitting at the Γ-point. The quantity $P$ is the momentum matrix element at the Γ-point. Using a 30-band $k \cdot p$ code,[55] we have studied the effect of distant bands on the effective mass $m_c$, and we find it is negligible. Therefore, the use of Eq. (57) is perfectly justified. Using experimental values for the effective mass and the band gap parameters from Aggarwal (Ref. 56), we find $P^2/2m_0 = 12.299$ eV at 30 K. The spin-orbit splitting is taken as $\Delta_0 = 0.287$ eV and as temperature-independent.[57] The temperature dependence of the effective mass is then determined by the temperature dependence of $E_0$ (Ref. 53) and $P$, which arises from fact that $P$ is expected to be inversely proportional to the lattice parameter.[58]

The valence band dispersion near the Γ-point of the Brillouin zone is characterized by the three independent Dresselhaus-Kip-Kittel (DKK) parameters $A$, $B$, and $|C|$ (Ref. 54). Expressions for these parameters that account for their temperature dependence have been given in Appendix B of Ref. 59. Because of the strong band warping, hole effective masses can only be defined by appropriate angular averages of the dispersion relations. Since in most expressions the masses appear in quantities involving densities of states, we obtain the heavy- and light-hole masses $m_{hh}^{3/2}$ and $m_{lh}^{3/2}$ as angular averages of the corresponding expressions in terms of $A$, $B$, and $|C|$.[59] This gives at 20 °C $m_{hh} = 0.352 m_0$ and $m_{lh} = 0.0385 m_0$.

For TAT calculations we need dispersion masses along the [001] direction. These are given by[58]

$$\frac{m_0}{m_{c,hh}} = -A + B \quad (58)$$

for heavy-holes and by

$$\frac{m_0}{m_{c,lh}} = -A - B \quad (59)$$

for light holes. This gives at 20 °C $m_{c,hh} = 0.403 m_0$ and $m_{c,lh} = 0.0829 m_0$. Our semiconductor equations, however, are based on a two-band model that does not distinguish between light- and heavy holes, so we compute the hole enhancement function as a weighted average of the heavy- and light-hole enhancement functions:

$$\Gamma_{pt} = \frac{m_{hh}^{3/2}}{m_{hh}^{3/2} + m_{lh}^{3/2}}\Gamma_{hh,t} \quad (60)$$
$$+ \frac{m_{lh}^{3/2}}{m_{hh}^{3/2} + m_{lh}^{3/2}}\Gamma_{lh,t}.$$

Here we have chosen weights approximately proportional to the corresponding densities of state. Given the dominance of heavy holes in this respect, the error incurred by making this ad hoc average assumption cannot be very large.

### 6. Strain

Ge films grown on Si or other substrates are usually tetragonally distorted due to the lattice mismatch between film and substrate. This mismatch is mostly relaxed in good-quality Ge-on-Si, but the residual levels of strain can still have a non-negligible effect on the I-V characteristics and their temperature dependence, since they affect the band gaps. We correct for the strain dependence of the band gaps using the expressions in Ref. 60. We use the same values of the deformation potentials given in this reference except that we take -$b_1$ = 2.88 eV and $b_2$ = 0 eV, consistent with the more recent experimental results of Liu and coworkers.[61] Note that the expressions in Ref. 60 represent an attempt to compute the shifts of individual bands relative to an absolute energy reference, whereas here we are interested in the band gap strain dependence. This leads to a simplification of the corresponding expressions.

The strain determined experimentally is only valid at the temperature it was measured. For simulations of the temperature dependence of the I-V curves, we need to consider the possibility of additional strain due the thermal expansivity mismatch between Ge and the substrate or buffer layer used. For Ge-on-Si this additional strain can be computed based on Eqs. (53) and (54) by assuming that the film is forced to track the substrate's thermal expansion. Strain also affects the effective masses, but we do not include such effects here. Similarly, the weights in the average computed in Eq. (60) should depend on strain because strain splits the light- and heavy-hole bands, but we also neglect this effect.

### 7. Bimolecular recombination

Bimolecular recombination plays a very limited role in all diodes examined here, but we included it for completeness. We computed the coefficient $B_{\text{bim}}$ using the theory for direct and indirect absorption discussed in Ref. 19. The results for Ge at 20 °C are $B_{\text{bim,dir}}/n_i^2 = 8.89 \times 10^{12}$ cm$^{-3}$s$^{-1}$ for the direct gap and $B_{\text{bim,ind}}/n_i^2 = 1.23 \times 10^{12}$ cm$^{-3}$s$^{-1}$ for the fundamental indirect gap.

**F.  Thermal, transport and electrical properties**

### 8. Thermal properties

Accurate values of the intrinsic carrier concentration $n_i$ and associated Fermi levels are important for matching theoretical and experimental I-V curves and modeling their temperature dependence. As noted in Ref. 62, there are non-neglible deviations between experimental and theoretical values of $n_i$ if one uses elementary textbook expressions based on the effective masses introduced above. This is due to significant non-parabolicity effects in Ge. Thus we compute the intrinsic carrier concentrations using the model described in detail in Ref. 19 by setting $n = p$ and solving for the common Fermi level. For these calculations we use strain-corrected values for the band gaps included. The only additional deformation potentials needed beyond those described in Ref. 60 are the ones corresponding to the third conduction band minimum along the $\Delta$-direction (which in Si is the absolute conduction band minimum). These deformation potentials are defined in the Appendix of Ref. 63, where recommended values for Si and Ge are given.

### 9. Mobilities

For the electron mobilities, we use the expressions proposed in Ref. 46 which were fit to experimental data as described there. For the hole mobilities, we use

$$\mu_p = \frac{\mu_{p0}(T)}{1 + \left(\frac{p}{A \times 10^{17}\mathrm{cm}^{-3}}\right)^{\alpha_0}}, \qquad (61)$$

which we fit to the experimental data from Trumbore and Tartaglia[64] using $\mu_{p0}(T) = 1.05 \times 10^9 \times T^{-2.33}$ cm$^2$/Vs from Ref. 65. The fit parameters are $A = 1.9349$ and $\alpha_0 = 0.4302$.

### 10. Recombination lifetimes

#### a. Bulk lifetimes

The parameters that appear in Eq. (42) must be fit to experimental data. We have attempted to fit results from several references[66-71] While there is no fully satisfactory fit of the experimental data with a simple analytical expression, we find reasonable agreement assuming a mid-gap defect and

$$\tau_n = \frac{\tau_{n0}}{1 + N/N_0}\left(\frac{300\ K}{T}\right)^{3/2} \qquad (62)$$

and

$$\tau_n = \frac{\tau_{p0}}{1 + N/N_0}\left(\frac{300\ K}{T}\right)^{3/2} \qquad (63)$$

with $\tau_{n0} = 5$ µs, $\tau_{p0} = 100$ µs, $N_0 = 8 \times 10^{15}$ cm$^{-3}$, and $N$ the dopant density. The temperature dependence is from Schenk.[72]

#### b. Dislocation lifetimes

Yamaguchi and coworkers derived an expression for the recombination lifetime associated with dislocations that was useful to understand the behavior of GaAs solar cells on Si.[73] However, we find that for the case of Ge-on-Si the Yamaguchi expression gives lifetimes that appear too short compared to experimental data. For example, for dislocation densities in the neighborhood of $10^7$ cm$^{-2}$, the expression predicts lifetimes near 100 ps, but the experimental values are in the µs range.[14] Unfortunately, the experimental data is scarce and large variations of more than one order of magnitude are observed for nominally similar dislocation concentrations, suggesting that the correlation between the number of recombination centers and the dislocation density may depend on the precise way dislocations are generated or annealed out. For this reason, attempts to fit general expressions to subsets of the available data seem of dubious general validity. We then take the opposite approach. We start with all existing information, nicely summarized by Son and coworkers,[14] and we add data from DiLello and coworkers.[11] When all these data are combined, a linear dependence between inverse recombination lifetimes and dislocation density cannot be ruled out, although some individual sets of data, such as the results from Eneman and coworkers,[74] are strongly superlinear. We then take the simplest approach and carry out a linear fit, the result being Eq. (51). It is important to point out that our models make it very difficult to distinguish between recombination/generation due to dislocations and contamination during growth. The fact that we fit dislocation densities similar to the measured values suggests that contamination effects are modest.

## APPENDIX B: SEMICONDUCTOR EQUATION COEFFICIENTS

We showed above that the discretized semiconductor equations can be all written as shown in Eq. (23). We then defined dimensionless quantities in Eqs. (30)-(33) to facilitate the numerical computations. Table B.1 shows the coefficients of Eq. (23) in terms of those dimensionless quantities.

:

| | $v$ | $u_n$ | $u_p$ |
|---|---|---|---|
| $\alpha_{j+1}$ | $-\dfrac{2\epsilon'_{j+\frac{1}{2}}}{(h'_j + h'_{j-1})h'_j}$ | $\dfrac{2D'_{n,j+\frac{1}{2}}}{(h'_j + h'_{j-1})h'_j} B(v_j - v_{j+1})\exp(v_j)$ | $-\dfrac{2D'_{p,j+\frac{1}{2}}}{(h'_j + h'_{j-1})h'_j} B(v_{j+1} - v_j)\exp(-v_j)$ |
| $\alpha_{j-1}$ | $-\dfrac{2\epsilon'_{j-\frac{1}{2}}}{(h'_j + h'_{j-1})h'_{j-1}}$ | $\dfrac{2D'_{n,j-\frac{1}{2}}}{(h'_j + h'_{j-1})h'_{j-1}} B(v_j - v_{j-1})\exp(v_j)$ | $-\dfrac{2D'_{p,j-\frac{1}{2}}}{(h'_j + h'_{j-1})h'_{j-1}} B(v_{j-1} - v_j)\exp(-v_j)$ |
| $\alpha_j$ | $\alpha_{j+1} + \alpha_{j-1}$ | | |
| $G_j$ | $\dfrac{n_i}{\bar{N}}\{u_{pj}\exp(-v_j) - u_{nj}\exp(v_j)\}$ | $u_{n,j}u_{p,j}(r'_j + B'_{j,\text{bim}})$ | $-u_{n,j}u_{p,j}(r'_j + B'_{j,\text{bim}})$ |
| $f_j$ | $\dfrac{(N_{dj} - N_{aj})}{\bar{N}}$ | $-(r'_j + B'_{j,\text{bim}}) - g'_j$ | $(r'_j + B'_{j,\text{bim}}) + g'_j$ |

**Table B1.** Coefficients of Eq. (23) for each of the independent functions $v$, $u_n$, and $u_p$. Notice that we have introduced a term $g'_j$ not defined in the main text. This term must be added when the diode is illuminated. Explicit expressions for $g'_j$ will be given elsewhere.

**REFERENCES**

1. M. T. Currie, S. B. Samavedam, T. A. Langdo, C. W. Leitz and E. A. Fitzgerald, Applied Physics Letters **72**, 1718-1721 (1998).

2. V. Sorianello, L. Colace, N. Armani, F. Rossi, C. Ferrari, L. Lazzarini and G. Assanto, Opt. Mater. Express **1** (5), 856-865 (2011).

3. K. H. Lee, S. Bao, Y. Lin, W. Li, P. Anantha, L. Zhang, Y. Wang, J. Michel, E. A. Fitzgerald and C. S. Tan, Journal of Materials Research **32** (21), 4025-4040 (2017).

4. L. Colace and G. Assanto, Photonics Journal, IEEE **1** (2), 69 (2009).

5. J. Michel, J. Liu and L. C. Kimerling, Nat Photon **4** (8), 527-534 (2010).

6. J. Wang and S. Lee, Sensors (Basel) **11** (1), 696-718 (2011).

7. X. Zhao, M. Moeen, M. S. Toprak, G. Wang, J. Luo, X. Ke, Z. Li, D. Liu, W. Wang, C. Zhao and H. H. Radamson, Journal of Materials Science: Materials in Electronics **31** (1), 18-25 (2019).

8. R. A. Soref, J. Kouvetakis, J. Menéndez, J. Tolle and V. R. D'Costa, Journal of Materials Research **22** (12), 3281-3291 (2007).

9. C. L. Senaratne, J. D. Gallagher, T. Aoki, J. Kouvetakis and J. Menéndez, Chemistry of Materials **26** (20), 6033-6041 (2014).

10. S. Wirths, D. Buca and S. Mantl, Progress in Crystal Growth and Characterization of Materials **62** (1), 1-39 (2016).

11. N. A. DiLello, D. K. Johnstone and J. L. Hoyt, Journal of Applied Physics **112** (5), 054506 (2012).

12. H. Chen, P. Verheyen, P. De Heyn, G. Lepage, J. De Coster, S. Balakrishnan, P. Absil, G. Roelkens and J. Van Campenhout, Journal of Applied Physics **119** (21) (2016).

13. E. Simoen, B. Hsu, G. Eneman, E. Rosseel, R. Loo, H. Arimura, N. Horiguchi, W. C. Wen, H. Nakashima, C. Claeys, A. Oliveira, P. Agopian and J. Martino, presented at the 2019 34th Symposium on Microelectronics Technology and Devices (SBMicro), 2019 (unpublished).

14. B. Son, Y. Lin, K. H. Lee, Q. Chen and C. S. Tan, Journal of Applied Physics **127** (20) (2020).

15. A. Pizzone, S. A. Srinivasan, P. Verheyen, G. Lepage, S. Balakrishnan and J. V. Campenhout, presented at the 2020 IEEE Photonics Conference (IPC), 2020 (unpublished).

16. X. Zhao, G. Wang, H. Lin, Y. Du, X. Luo, Z. Kong, J. Su, J. Li, W. Xiong, Y. Miao, H. Li, G. Guo and H. H. Radamson, Nanomaterials (Basel) **11** (5) (2021).

17. C. Xu, R. T. Beeler, L. Jiang, G. Grzybowski, A. V. G. Chizmeshya, J. Menéndez and J. Kouvetakis, Semiconductor Science and Technology **28** (10), 105001 (2013).

18. S. M. Sze, *Physics of Semiconductor Devices*. (Wiley, New York, 1981).

19. J. Menéndez, C. D. Poweleit and S. E. Tilton, Physical Review B **101** (19), 195204 (2020).

20. E. Rosencher and B. Vinter, *Optoelectronics*. (Cambridge University Press, Cambridge, 2002).

21. C. E. Korman and I. D. Mayergoyz, Journal of Applied Physics **68** (3), 1324-1334 (1990).

22. D. L. Scharfetter and H. K. Gummel, IEEE Transactions on Electron Devices **16** (1), 64-77 (1969).

23. C. M. Snowden, *Semiconductor device modelling*. (Peter Peregrinus, 1988).

24. H. Sundqvist and G. Veronis, Tellus A: Dynamic Meteorology and Oceanography **22** (1) (1970).

25. I. D. Mayergoyz, Journal of Applied Physics **59** (1), 195-199 (1986).

26. W. W. Keller, Journal of Applied Physics **61** (11), 5189-5190 (1987).

27. T. I. Seidman and S. C. Choo, Solid-State Electronics **15** (11), 1229-1235 (1972).

28. W. H. Press, S. A. Teukolsky, W. T. Vetterling and B. P. Flannery, *Numerical Recipes in C: The Art of Scientific Computing*, Second Edition ed. (Cambridge University Press, New York, 1992).

29. G. A. M. Hurkx, D. B. M. Klaassen and M. P. G. Knuvers, IEEE Transactions on Electron Devices **39** (2), 331-338 (1992).

30. W. Takeuchi, T. Asano, Y. Inuzuka, M. Sakashita, O. Nakatsuka and S. Zaima, ECS Journal of Solid State Science and Technology **5** (4), P3082-P3086 (2015).

31. S. Gupta, E. Simoen, R. Loo, Y. Shimura, C. Porret, F. Gencarelli, K. Paredis, H. Bender, J. Lauwaert, H. Vrielinck and M. Heyns, Applied Physics Letters **113** (2) (2018).

32. B. Wang, Z. Q. Fang, B. Claflin, D. Look, J. Kouvetakis and Y. K. Yeo, Thin Solid Films **654**, 77-84 (2018).

33. A. Czerwinski, E. Simoen, A. Poyai and C. Claeys, Journal of Applied Physics **94** (2), 1218-1221 (2003).

34. L. Colace, G. Masini, F. Galluzzi, G. Assanto, G. Capellini, L. D. Gaspare, E. Palange and F. Evangelisti, Applied Physics Letters **72** (24), 3175-3177 (1998).

35. L. Colace, P. Ferrara, G. Assanto, D. Fulgoni and L. Nash, Photonics Technology Letters, IEEE **19** (22), 1813 (2007).

36. M. Balbi, V. Sorianello, L. Colace and G. Assanto, Physica E: Low-dimensional Systems and Nanostructures **41** (6), 1086-1089 (2009).

37. Y.-Y. Fang, J. Tolle, J. Tice, A. V. G. Chizmeshya, J. Kouvetakis, V. R. D'Costa and J. Menéndez, Chemistry of Materials **19** (24), 5910-5925 (2007).

38. G. Grzybowski, L. Jiang, R. T. Beeler, T. Watkins, A. V. G. Chizmeshya, C. Xu, J. Menéndez and J. Kouvetakis, Chemistry of Materials **24** (9), 1619-1628 (2012).

39. G. Grzybowski, A. V. G. Chizmeshya, C. Senaratne, J. Menendez and J. Kouvetakis, Journal of Materials Chemistry C **1** (34), 5223 (2013).

40. C. Xu, J. D. Gallagher, P. M. Wallace, C. L. Senaratne, P. Sims, J. Menéndez and J. Kouvetakis, Semiconductor Science and Technology **30** (10), 105028 (2015).

41. A. Zhang, M. A. Mircovich, D. A. Ringwala, C. D. Poweleit, M. A. Roldan, J. Menéndez and J. Kouvetakis, Journal of Materials Chemistry C **10** (36), 13107-13116 (2022).

42. C. Xu, C. L. Senaratne, J. Kouvetakis and J. Menéndez, Applied Physics Letters **105** (23), 232103 (2014).

43. C. Xu, C. L. Senaratne, J. Kouvetakis and J. Menéndez, Physical Review B **93** (4), 041201 (2016).

44. C. Xu, N. S. Fernando, S. Zollner, J. Kouvetakis and J. Menéndez, Physical Review Letters **118** (26), 267402 (2017).

45. C. Xu, J. Kouvetakis and J. Menéndez, Journal of Applied Physics **125** (8), 085704 (2019).

46. J. Menéndez, C. Xu and J. Kouvetakis, Materials Science in Semiconductor Processing **164**, 107596 (2023).

47. C. Xu, T. Hu, D. A. Ringwala, J. Menéndez and J. Kouvetakis, Journal of Vacuum Science & Technology A **39** (6), 063411 (2021).

48. V. R. D'Costa, Y. Fang, J. Mathews, R. Roucka, J. Tolle, J. Menendez and J. Kouvetakis, Semiconductor Science and Technology **24** (11), 115006 (2009).

49. C. Xu, P. M. Wallace, D. A. Ringwala, J. Menendez and J. Kouvetakis, ACS Appl Mater Interfaces **10** (43), 37198-37206 (2018).

50. R. Roucka, Y. Y. Fang, J. Kouvetakis, A. V. G. Chizmeshya and J. Menéndez, Physical Review B **81** (24), 245214 (2010).

51. H. Ibach, physica status solidi (b) **33** (1), 257-265 (1969).

52. Y. Ma and J. S. Tse, Solid State Communications **143** (3), 161 (2007).

53. C. Emminger, N. S. Samarasingha, M. Rivero Arias, F. Abadizaman, J. Menéndez and S. Zollner, Journal of Applied Physics **131** (16), 165701 (2022).

54. G. Dresselhaus, A. Kip and C. Kittel, Physical Review **98** (2), 368-384 (1955).

55. D. Rideau, M. Feraille, L. Ciampolini, M. Minondo, C. Tavernier, H. Jaouen and A. Ghetti, Physical Review B **74** (19), 195208 (2006).

56. R. L. Aggarwal, Physical Review B **2** (2), 446-458 (1970).

57. D. E. Aspnes, Physical Review B (Condensed Matter and Materials Physics) **12** (6), 2297-2310 (1975).

58. P. Y. Yu and M. Cardona, *Fundamentals of Semiconductors: Physics and Materials Properties*. (Springer-Verlag, Berlin, 1996).

59. J. Menéndez, D. J. Lockwood, J. C. Zwinkels and M. Noël, Physical Review B **98** (16), 165207 (2018).

60. J. Menendez and J. Kouvetakis, Applied Physics Letters **85** (7), 1175-1177 (2004).

61. J. Liu, D. D. Cannon, K. Wada, Y. Ishikawa, D. T. Danielson, S. Jongthammanurak, J. Michel and L. Kimerling, Physical Review B (Condensed Matter and Materials Physics) **70** (15), 155309 (2004).

62. J. Menéndez, P. M. Wallace, C. Xu, C. L. Senaratne, J. D. Gallagher and J. Kouvetakis, Materials Today: Proceedings **14**, 38-42 (2019).

63. J. Teherani, W. Chern, D. Antoniadis, J. Hoyt, L. Ruiz, C. Poweleit and J. Menéndez, Physical Review B **85** (20), 205308 (2012).

64. F. A. Trumbore and A. A. Tartaglia, Journal of Applied Physics **29** (10), 1511-1511 (1958).

65. O. Madelung, in *Landolt-Börstein: Numerical Data and Functional Relationships in Science and Technology*, edited by K. H. Hellwege (Springer-Verlag, Berlin, 1985), Vol. 17a.

66. E. Gaubas, M. Bauža, A. Uleckas and J. Vanhellemont, Materials Science in Semiconductor Processing **9** (4-5), 781-787 (2006).

67. E. Gaubas and J. Vanhellemont, Applied Physics Letters **89** (14), 142106 (2006).

68. E. Gaubas and J. Vanhellemont, Journal of The Electrochemical Society **154** (3) (2007).

69. E. Gaubas, J. Vanhellemont, E. Simoen, I. Romandic, W. Geens and P. Clauws, Physica B: Condensed Matter **401-402**, 222-225 (2007).

70. B. P. Swain, H. Takato, Z. Liu and I. Sakata, Solar Energy Materials and Solar Cells **95** (1), 84-88 (2011).

71. B. P. Swain, H. Takato and I. Sakata, Japanese Journal of Applied Physics **50** (7), 071302 (2011).

72. A. Schenk, Solid-State Electronics **35** (11), 1585-1596 (1992).

73. M. Yamaguchi, A. Yamamoto and Y. Itoh, Journal of Applied Physics **59** (5), 1751-1753 (1986).

74. G. Eneman, M. Bargallo Gonzalez, G. Hellings, B. De Jaeger, G. Wang, J. Mitard, K. DeMeyer, C. Claeys, M. Meuris, M. Heyns, T. Hoffmann and E. Simoen, ECS Transactions **28** (5), 143 (2010).